\pdfoutput=1
\PassOptionsToPackage{
pdfencoding=auto,
pdfnewwindow=true,
pdfusetitle=true,
bookmarks=true,
bookmarksnumbered=true,
bookmarksopen=true,
pdfpagemode=UseThumbs,
bookmarksopenlevel=1,
pdfpagelabels=false,
breaklinks=true,
}{hyperref}
\documentclass[10pt,showpacs,letterpaper,aps,twocolumn,nofootinbib,pra]{revtex4-2} 
\usepackage{amsthm,amsmath,amssymb,graphicx,comment} 
\usepackage{bbold}
\usepackage[none]{hyphenat}

\usepackage{microtype}
\microtypecontext{spacing=nonfrench}
\microtypesetup{
expansion={true,nocompatibility},
protrusion={true,nocompatibility},
activate={true,nocompatibility},
tracking=true,
kerning=true,
spacing={true}
}

\usepackage[all=normal,floats=tight,mathspacing=tight,wordspacing=tight,paragraphs=normal,tracking=tight,charwidths=tight,mathdisplays=normal,sections=normal,margins=normal]{savetrees}

\everypar=\expandafter{\the\everypar\loosness=-1 }
\usepackage{units}
\usepackage{multirow}

\usepackage[usenames,dvipsnames,table]{xcolor}
\definecolor{purple}{RGB}{128,0,128}
\definecolor{ultramarine}{RGB}{63, 0, 255}
\definecolor{medblue}{RGB}{0, 0, 100}
\definecolor{panblue}{RGB}{0,24,150}
\definecolor{carmine}{RGB}{150, 0, 24}
\definecolor{gray}{RGB}{150, 150, 150}
\definecolor{darkred}{RGB}{200, 0, 0}
\definecolor{darkgreen}{RGB}{0, 80, 0}
\definecolor{darkblue}{RGB}{0, 0, 200}

\definecolor{nred}{rgb}{0.9,0.1,0.1}
\definecolor{nblack}{rgb}{0,0,0}
\definecolor{nblue}{rgb}{0.2,0.2,0.8}
\definecolor{ngreen}{rgb}{0.2,0.6,0.2}

\usepackage{hyperref}
\hypersetup{colorlinks, breaklinks, linkcolor=darkred, urlcolor=darkblue, citecolor=darkgreen}

\usepackage{bm}

\newcommand{\beq}{\begin{equation}}
\newcommand{\eeq}{\end{equation}}
\newcommand{\bea}{\begin{align}}
\newcommand{\eea}{\end{align}}

\newtheoremstyle{formltn}%
{}%
{}%
{\itshape}%
{}%
{\bfseries}%
{ of the existence of a universally noncontextual model:}%
{\parindent}%
{\thmname{#1} \thmnumber{{#2}}\thmnote{\normalfont#3}}%
\theoremstyle{formltn}
\newtheorem{formulation}{Formulation}
\newtheoremstyle{thepolytope}%
{}%
{}%
{\itshape}%
{}%
{\bfseries}%
{ of the noncontextual measurement-assignment polytope:}%
{\parindent}%
{\thmname{#1} \thmnumber{{#2}}\thmnote{\normalfont#3}}%
\theoremstyle{thepolytope}
\newtheorem{thepolytope}{Characterization}

\usepackage{etoolbox}

\AtBeginEnvironment{formulation}{
\begin{subequations}
}
\AtEndEnvironment{formulation}{
\end{subequations}
}

\begin{document}
\title{All the noncontextuality inequalities for arbitrary prepare-and-measure experiments with respect to any fixed sets of operational equivalences}
\author{David Schmid, Robert W. Spekkens, and Elie Wolfe}
\affiliation{Perimeter Institute for Theoretical Physics, 31 Caroline Street North, Waterloo, Ontario Canada N2L 2Y5}
\email{dschmid@perimeterinstitute.ca}
\begin{abstract}
Within the framework of generalized noncontextuality, we introduce a general technique for systematically deriving noncontextuality inequalities for any experiment involving finitely many preparations and finitely many measurements, each of which has a finite number of outcomes. Given any fixed sets of operational equivalences among the preparations and among the measurements as input, the algorithm returns a set of noncontextuality inequalities whose satisfaction is necessary and sufficient for a set of operational data to admit of a noncontextual model. 
Additionally, we show that the space of noncontextual data tables always defines a polytope. 
Finally, we provide a computationally efficient means for testing whether any set of numerical data admits of a noncontextual model, with respect to any fixed operational equivalences. Together, these techniques provide complete methods for characterizing arbitrary noncontextuality scenarios, both in theory and in practice. 
Because a {\em quantum} prepare-and-measure experiment admits of a noncontextual model if and only if it admits of a positive quasiprobability representation, our techniques also determine the necessary and sufficient conditions for the existence of such a representation. 
\end{abstract}
\maketitle

\section{Introduction}

Proofs of the failure of noncontextuality place strong constraints on our understanding of nature. For example, the argument by Kochen and Specker~\cite{KS} showed that quantum theory cannot be explained by an ontological model in which every projective measurement has its outcome determined by hidden variables, independently of what other measurements it is implemented jointly with (that is, independently of context). 

A generalized notion of noncontextuality was defined in Ref.~\cite{gencontext}. Heuristically, it asserts that an ontological model of some operational theory is noncontextual if and only if laboratory operations which cannot be operationally distinguished are represented identically in the model. It was shown that the operational predictions of quantum theory are inconsistent with the existence of such a model. Furthermore, this generalized notion of noncontextuality (which we henceforth refer to simply as ``noncontextuality'') applies to arbitrary operational theories---not just quantum theory---so one can ask whether \textit{any} given set of data describing possible statistics for a prepare-and-measure experiment admits of a noncontextual model. We will refer to such sets of data as \textit{data tables}, since one can organize the data into a table giving the probability for each pairing of a preparation with a measurement outcome.

Because the generalized notion of noncontextuality of Ref.~\cite{gencontext} presumes neither the correctness of quantum theory nor that the preparations and measurements are noiseless, its operational consequences can be tested directly with real experimental data\footnote{Other articles, e.g. Refs.~\cite{cab1,cab4,cab5}, have instead sought to make the Kochen-Specker notion of noncontextuality applicable to arbitrary operational theories and thereby to derive experimental tests of this notion. These proposals have been contrasted with those based on generalized noncontextuality and criticized in Refs.~\cite{determinism, unwarranted,statei}.}. These tests take the form of \textit{noncontextuality inequalities}, whose violation by a data set witnesses the impossibility of explaining that data with a noncontextual ontological model. One such inequality has been violated with high confidence in a recent experiment, demonstrating that nature does not admit of a noncontextual model~\cite{unwarranted}. 
 
This article presents a systematic method for characterizing, for any prepare-and-measure experiment, all the data tables consistent with the principle of noncontextuality given any fixed set of operational equivalences among the preparations and any fixed set of operational equivalences among the measurements. The inputs to our algorithm are these two sets of operational equivalences, and the output is a set of inequalities on the elements of the data table. The satisfaction of these inequalities is necessary and sufficient for the data table to admit of a noncontextual model with respect to the specified operational equivalences. Our method shows that the space of noncontextual data tables for any prepare-and-measure scenario defines a polytope, which we term the \textit{generalized-noncontextual polytope}, in analogy with the \textit{local polytope} of a Bell scenario~\cite{Bellreview}.

In addition, this article provides a systematic and efficient method for deciding if any given data table, specified numerically, admits of a noncontextual model, given any fixed set of operational equivalences.

Foundationally, the failure of noncontextuality plays a key role as a notion of nonclassicality which has broad applicability, and which subsumes other notions such as the negativity of quasiprobability representations~\cite{negativity}, the generation of anomalous weak values~\cite{AWV}, and violations of local causality~\cite{gencontext}. As such, our method for deriving inequalities allows one to quantitatively identify the boundary between the classical and the nonclassical in arbitrary prepare-and-measure scenarios. Furthermore, it allows one to directly generate noise-robust noncontextuality inequalities from arbitrary proofs of the failure of noncontextuality in quantum theory~\cite{twothms}, 
thereby allowing one to find necessary and sufficient conditions for noncontextuality in scenarios where only necessary conditions were previously known~\cite{POM,unwarranted,stated,statei,Ravi}.

Practically, quantifying the boundary between classical and nonclassical is important for identifying tasks which admit of a quantum advantage. For example, the failure of noncontextuality has been shown to be a resource in various tasks, providing advantages for cryptography~\cite{POM,RAC,RAC2,deba1} and computation~\cite{magic,comp1,comp2}. 
If a given protocol for quantum cryptography/communication or a given quantum circuit can be cast into the form of a prepare-and-measure experiment, then one can apply our methods to determine whether the given protocol or circuit admits of a noncontextual model. %
By this sort of study, one might identify new quantum information-processing tasks for which contextuality is a resource. Our methods should also prove useful for the benchmarking of real quantum devices, just as Bell inequality violations allow one to certify randomness~\cite{randomness} and achieve key distribution~\cite{DIQKD,DIQKD2} in a device-independent fashion.

Identifying operational equivalences among laboratory procedures is the first step in deriving observable consequences of generalized noncontextuality. The particular operational equivalences considered, in turn, determine the set of %
noncontextuality inequalities which follow. As such, it is important to understand where the operational equivalences come from in the first place. Historically, most operational equivalences for generalized noncontextuality have originated in quantum no-go theorems for noncontextuality~\cite{peresmermin,POM,unwarranted,stated,statei,deba2}, or in a frustrated network that can be used to build a quantum no-go theorem~\cite{LSW,specker}. These quantum no-go theorems are typically designed to be as simple as possible or to highlight specific structures within quantum theory.
One then extrapolates to an operational setting by identifying a set of (imperfect, noisy) operational procedures that satisfy the precise operational equivalence relations that held among the idealized (perfect, noiseless) quantum procedures. By considering procedures that are convex mixtures of the ones actually realized in an experiment, it is possible to satisfy these particular operational equivalences exactly~\cite{unwarranted}. %
To approximate what one would expect quantumly, it is important that the experiment be engineered to target the particular operational equivalences in question. However, there is a different approach that one can adopt: rather than trying to engineer one's experiment to accommodate some specific sets of operational equivalences, 
one can instead take a set of data from an arbitrary experiment, and then simply determine the operational equivalences that it naturally exhibits~\footnote{\label{foot1}A procedure for finding complete sets of operational equivalences was demonstrated in a special case in Ref.~\cite{robust}. A more general method is forthcoming.}. The only requirement that such experiments must fulfill is that the set of preparations and the set of measurements are tomographically complete (as discussed in depth in Refs.~\cite{robust,unwarranted}).  Operational equivalences extracted directly from experimental data will not generally not be as simple in structure as those exhibited in quantum no-go theorems (for instance, the latter often involve {\em uniform} mixtures and are often symmetric under certain permutations of the procedures).  However, the techniques we provide in this paper apply just as well to them. Further, because one is not confined to testing a predetermined set of operational equivalences, one can test for failures of noncontextuality in generic experiments, rather than just those which target laboratory procedures motivated by some specific no-go theorem.

Our method for finding the generalized-noncontextual polytope is comprised of two distinct computational tasks. The first task is to catalog all extremal solutions which satisfy some initial set of linear constraints; i.e., it is an instance of the vertex enumeration problem. That catalog allows us to formulate a condition for membership in the generalized-noncontextual polytope in terms of an existential quantifier. The second task, then, is quantifier elimination, and requires eliminating variables for a system of linear inequalities. As elaborated herein, a variety of standard algorithms are readily available for efficiently solving both of these tasks. Moreover, vertex enumeration and quantifier elimination algorithms are already widely used in quantum foundations\footnote{Familiar applications of vertex enumeration and quantifier elimination---albeit not always referred to by these names---include deriving standard Bell inequalities~\cite{Collins2004,Brunner2008,bancal2010symmetric,budroni}, deriving entropic inequalities for generalized correlation scenarios~\cite{Fritz2013Marginal,chavesluftgross,Chaves2015InfoTheo}, and many others~\cite{Anirudh,peresmermin,inflation,statei,NCFraction,scarani2011extremaltripartite}. For instance, Bell inequalities are defined as the convex hull of all local strategies; i.e. they precisely characterize the region in probability-space spanned by deterministic strategies~\cite{Bellreview}. This relates to vertex enumeration, because the convex hull problem (finding inequalities given extreme points) is \emph{equivalent} to vertex enumeration (finding extreme points given inequalities) from an algorithmic perspective (see for example Ref.~\cite[App. A]{peresmermin}).}.

Finally, we note that a prepare-and-measure experiment that is consistent with quantum theory
(i.e., a set of quantum preparations and a set of quantum measurements) admits of a positive quasiprobability representation if and only if it admits of a noncontextual ontological model~\cite{negativity}. We can therefore apply our technique to solve the problem of determining whether or not a given set of quantum preparations and measurements admits of a positive quasiprobability representation~\cite{FerrieEmerson,negativity,posquasi}.  It suffices to determine all of the operational equivalences holding among the preparations and among the measurements (see Footnote~\ref{foot1}),
and then determine whether the data table defined by the prepare-and-measure experiment violates any noncontextuality inequality that is implied by these operational equivalences.

\section{Operational and ontological preliminaries}

An operational theory specifies a set of laboratory procedures, such as preparations and measurements, as well as a prescription for finding the probability distribution $p(m|M,P)$ over outcomes $m$ of any given measurement $M$ when implemented following any given preparation $P$. 

Two preparations $P_1$ and $P_2$ are termed \textit{operationally equivalent} if they generate the same statistics for all possible measurements:
\beq \label{OEp}
\forall M: p(m|M,P_1) = p(m|M,P_2).
\eeq
We denote this operational equivalence relation by ${P_1 \simeq P_2}$.%

Similarly, two measurement procedures $M_1$ and $M_2$ are termed \textit{operationally equivalent} if they generate the same probability distribution over their outcomes (denoted $m_1$ and $m_2$) for all possible preparations:
\beq \label{OEm}
\forall P: p(m_1|M_1,P) = p(m_2|M_2,P).
\eeq
We denote this operational equivalence relation by ${M_1 \simeq M_2}$.%

We will refer to the event of a measurement $M$ yielding an outcome $m$ as a \textit{measurement effect}, denoted $[m|M]$.

If one samples the choice of laboratory procedure $O_i$ from some probability distribution $p_i$ and then forgets the value of $i$, we introduce the shorthand notation $\sum_i p_i O_i$ for the effective procedure so defined.
The procedures here could be preparations or measurement effects.

An ontological model attempts to explain the probabilities $p(m|M,P)$ in an operational theory via a set $\Lambda$ of ontic states. An ontic state $\lambda \in \Lambda$ specifies all the physical properties of the system, and causally mediates correlations between the preparation and the measurement. For every laboratory preparation $P$, the model specifies a probability distribution $\mu_P(\lambda)$, where
\begin{align} 
\forall \lambda: \quad  \mu_P(\lambda) \ge 0, \label{prob1} \\ 
\int_{\Lambda} d\lambda \, \mu_P(\lambda) = 1 \label{norm1}.
\end{align} 
Whenever preparation $P$ is implemented, the ontic state $\lambda$ is sampled from an associated probability distribution $\mu_P(\lambda)$.
Every measurement $M$ generates an outcome $m$ as a probabilistic function of the ontic state according to some fixed \textit{response function} $\xi_{m|M}(\lambda)$, where
\begin{alignat}{2}
\forall \lambda, m:&& \quad \xi_{m|M}(\lambda) \ge 0, \label{m1} \\ \label{m2}
\forall \lambda:&& \quad \sum_m \xi_{m|M}(\lambda) = 1 .
\end{alignat}
For the ontological model to reproduce an operational theory's empirical predictions, one requires that 
\begin{equation} \label{probs}
\forall m,M,P: \quad p(m | M,P) = \int_{\Lambda}  \xi_{m|M}(\lambda) \mu_P(\lambda)d\lambda .
\end{equation}

Finally, note that an effective laboratory procedure $\sum_i p_i O_i$ is represented in an ontological model by the corresponding convex mixture of the ontological representations of the individual operations $O_i$ (see Eq.~(7) of~\cite{negativity} and the surrounding discussion).

An ontological model which respects the principle of \textit{preparation noncontextuality}~\cite{gencontext} is one in which two operationally equivalent preparations are represented by the same distribution over ontic states; that is,
\begin{align} \label{PNC}
P_1 \simeq  P_2 & \ \ \text{implies that} \\ \nonumber
&\forall \lambda: \quad \mu_{P_1}(\lambda) = \mu_{P_2}(\lambda).
\end{align}
An ontological model which respects the principle of \textit{measurement noncontextuality}~\cite{gencontext} is one in which two operationally equivalent measurement effects are represented by the same response function; that is,
\begin{align} \label{MNC}
[m_1|M_1] \simeq  [m_2|M_2] & \ \ \text{implies that} \\ \nonumber
&\forall \lambda: \quad \xi_{m_1|M_1}(\lambda) = \xi_{m_2|M_2}(\lambda).
\end{align}
In this paper, the term noncontextuality refers to \textit{universal noncontextuality}~\cite{gencontext}, which posits noncontextuality for all procedures, including preparations and measurements.

\section{Problem setup}

The scenario we are considering has a set $\{P_1,P_2,...,P_g\}$ of $g$ preparations, a set $\{M_1,M_2,...,M_l \}$ of $l$ measurements, a set of $d$ outcomes $\{1,2,...,d \}$ for each measurement, a set of operational equivalences among the preparations, denoted $\mathcal{OE}_P$, and a set of operational equivalences among the measurements, denoted $\mathcal{OE}_M$. There are no restrictions on any of these sets, beyond the fact that they must be finite, as they are in any real experiment\footnote{In any real experiment with continuous variable systems, one must coarse-grain the outcomes to a finite set to obtain nonzero probabilities of any given event. There is also a nuance concerning experiments with a finite number of preparations, measurements and outcomes: the full set of operational equivalences among these might be infinite, but one can always find a finite generating set of operational equivalences whose implications for noncontextual data tables are equivalent to the implications of the full infinite set. %
See Footnote~\ref{foot1}.}. Furthermore, we have not presumed that one knows anything about the preparations and measurements, beyond the fact that they can be performed repeatedly so as to gather statistics. Without loss of generality, we treat each measurement as having exactly $d$ outcomes for some sufficiently large value of $d$ (since any measurement with $d^* < d$ outcomes can be redefined to have $d$ outcomes, $d-d^*$ of which never occur).

The input to our algorithm is a specification of the operational equivalences $\mathcal{OE}_P$ and $\mathcal{OE}_M$, and the desired output is a set of inequalities such that a data table $\{ p(m|M_i,P_j) \}_{i,j,m}$ admits a universally noncontextual model if and only if all the inequalities are satisfied. Although it is not obvious from the definition of universal noncontextuality, we will find that the final inequalities will be linear in the probabilities.

Generically, each of the operational equivalences $s~\in~\mathcal{OE}_P$ is of the form
\beq
\sum_j \alpha^{s}_{P_j} P_j \simeq \sum_{j'}  \beta^{s}_{P_{j'}} P_{j'}
\eeq
for some sets of convex weights $\{ \alpha^{s}_{P_j} \}_j$ and $\{ \beta^{s}_{P_{j'}} \}_{j'}$ (a set of convex weights is a list of nonnegative real numbers which sum to one).
Hence, the principle of preparation noncontextuality, Eq.~\eqref{PNC}, implies that the same functional relationships must hold among the ontological representations of the preparations; in other words,
\beq \label{Pconstraints}
\forall \lambda: \quad \sum_j  \alpha^{s}_{P_j} \mu_{P_j}(\lambda) = \sum_{j'}  \beta^{s}_{P_{j'}} \mu_{P_{j'}}(\lambda).
\eeq
Similarly, each of the operational equivalences $r \in \mathcal{OE}_M$ is of the form
\beq
\sum_{i,m} \alpha^{r}_{m|M_i} [m|M_i] \simeq \sum_{i',m'} \beta^{r}_{m'|M_{i'}} [m'|M_{i'}]
\eeq
for some sets of convex weights $\{ \alpha^{r}_{m|M_i} \}_{i,m}$ and $\{ \beta^{r}_{m'|M_{i'}} \}_{i',m'}$.
Hence, the principle of measurement noncontextuality, Eq.~\eqref{MNC}, implies that the same functional relationships must hold among the ontological representations of the effects; in other words,
\beq \label{m3}
\forall \lambda: \sum_{i,m} \alpha^{r}_{m|M_i} \xi_{m|M_i}(\lambda) = \sum_{i',m'}  \beta^{r}_{m'|M_{i'}} \xi_{m'|M_{i'}}(\lambda).
\eeq

The question of whether a data table admits a universally noncontextual model, then, may be compactly summarized as follows:
\setlength{\belowdisplayskip}{0pt plus 0pt}
\setlength{\abovedisplayskip}{0pt plus 0pt}
\setlength\abovedisplayshortskip{0pt}
\setlength\belowdisplayshortskip{0pt}
\begin{samepage}
\begin{formulation}\label{c1}
A universally noncontextual model for a data table $\{ p(m|M_i,P_j) \}_{i,j,m}$ exists (with respect to the sets of operational equivalences  $\mathcal{OE}_P$ and $\mathcal{OE}_M$)
 if and only if 
\vspace{1.5pt}\begin{align*}
\exists \Lambda,
\exists \{ \mu_{P_j}(\lambda)\}_{j,\lambda}, \{ \xi_{m|M_i}(\lambda)\}_{i,m,\lambda}& \text{  such that: } 
\end{align*}\vspace{2pt}\begin{alignat}{2}
\forall \lambda,i,m:&& \quad  \xi_{m|M_i}(\lambda) &\ge 0,  \label{cf}\\ 
\forall \lambda,i:&& \quad \sum_m \xi_{m|M_i}(\lambda) &= 1, \label{cs} \\ 
\hspace{-2ex}\forall \lambda,r  :&& \:\: \sum_{i,m} (\alpha^{r}_{m|M_i}\!-\beta^{r}_{m|M_i}) \xi_{m|M_i}(\lambda) &= 0, \label{ct} \\
\forall \lambda, j:&& \quad \mu_{P_j}(\lambda) &\ge 0, \\ 
\forall j:&& \quad \int_{\lambda} \mu_{P_j}(\lambda) &= 1, \\ 
\forall \lambda, s :&& \quad \sum_j  (\alpha^{s}_{P_j}-\beta^{s}_{P_j}) \mu_{P_j}(\lambda) &= 0, 
\end{alignat}\beq
\forall i,j,m:\quad\int_{\Lambda} \xi_{m|M_i}(\lambda) \mu_{P_j}(\lambda) d\lambda =  p(m | M_i,P_j). \label{cl}
\eeq
\end{formulation}
\end{samepage}
\normalsize
\noindent 
Eqs.~\eqref{cf} to ~\eqref{cl} represent, respectively: positivity of the response functions (Eq.~\eqref{prob1}); normalization of the response functions (Eq.~\eqref{norm1}), the consequences of noncontextuality implied by the operational equivalences in $\mathcal{OE}_M$ (Eq.~\eqref{m3}); positivity of the distributions associated with the preparations (Eq.~\eqref{m1}); normalization  of the distributions associated with the preparations (Eq.~\eqref{m2}); the consequences of noncontextuality implied by the operational equivalences in $\mathcal{OE}_P$ (Eq.~\eqref{Pconstraints}); and the expression for the probabilities in the data table in terms of the ontological model (Eq.~\eqref{probs}).

There are two key obstacles to deriving constraints directly on $\{ p(m|M_i,P_j) \}_{i,j,m}$ from the implicit constraints imposed by Eqs.~\eqref{cf}-\eqref{cl}. First, the ontic state space $\Lambda$ is unknown and possibly of unbounded cardinality, so that it is not obvious \textit{a priori}  whether there is an algorithm to solve the problem. Second, even if the number of ontic states were known to be finite, so that the problem could in principle be solved by quantifier elimination methods, the probabilities in Eq.~\eqref{cl} are nonlinear in the unknown parameters $\{\xi_{m|M_i}(\lambda)\}_{i,m,\lambda}$ and $\{\mu_{P_j}(\lambda)\}_{j,\lambda}$ appearing in the quantifiers; hence, the problem would be one of nonlinear quantifier elimination, which is computationally difficult. We overcome both of these problems by leveraging the convex structure of the space of response functions: we find the finite set of convexly-extremal noncontextual assignments to the measurements, identify the set of ontic states with it, and then parametrize the distributions corresponding to the preparations in terms of their probability assignments to these ontic states. Thus, the unknown parameters form a finite set, and furthermore the operational probabilities are linear in these parameters. 

Finally, we perform linear quantifier elimination to obtain constraints on the operational probabilities alone.

\section{Characterizing the generalized-noncontextual polytope} \label{gncpoly}

\subsection{Enumerating the convexly-extremal noncontextual measurement assignments} \label{mncpoly}

No matter what the form or size of the ontic state space $\Lambda$, a measurement-noncontextual assignment of probabilities to all $d$ outcomes of all $l$ measurements, for a particular ontic state $\lambda^*$, is an $(ld)$-component vector ${\bm{\xi}(\lambda^*)\!\equiv\!\Big(\! \xi_{1|M_1}(\lambda^*),\!...,\xi_{d|M_1}(\lambda^*), \xi_{1|M_2}(\lambda^*),\!..., \xi_{d|M_l}(\lambda^*)   \!\Big)}$ subject to the constraints of Eqs.~\eqref{cf}-\eqref{ct}. We call such an $(ld)$-component vector a \textit{noncontextual measurement assignment}. The set of all such assignments defines a polytope: 
\begin{samepage}\begin{thepolytope}\label{measurementpolytope}
The $(ld)$-component vector $\bm{\xi}(\lambda^*)$ lies inside the noncontextual measurement-assignment polytope if and only if
\begin{subequations}
\begin{align}
\forall i,m: & \quad  \xi_{m|M_i}(\lambda^*) \ge 0\,,  \\[4pt] 
\forall i: & \quad \sum_m \xi_{m|M_i}(\lambda^*) = 1\,,\\
\forall r: & \quad \sum_{i,m} (\alpha^{r}_{m|M_i}-\beta^{r}_{m|M_i}) \xi_{m|M_i}(\lambda^*) = 0\,.
\end{align}
\end{subequations}
\end{thepolytope}\end{samepage}
In what follows, it is critical to characterize this polytope by its vertices rather than its facets. The vertices are the convexly-extremal noncontextual measurement assignments. (Note that if there are {\em no} operational equivalences among the measurements, then these extremal assignments are deterministic, that is, all of the elements of the vector have value 0 or 1.) In general, to find the vertices of a polytope that is given in terms of its facet inequalities, one must solve the \textit{vertex enumeration problem}~\cite{Avis2000lrs,bremner_symmetries_2009,Zolotykh2012,avis_convexhull_2015,panda_2015}. Many excellent software packages are freely available for vertex enumeration\footnote{Dedicated software for performing vertex enumeration includes 
\texttt{traf} from \href[pdfnewwindow]{http://porta.zib.de}{\texttt{PORTA}}, 
\texttt{skeleton64f} from \href[pdfnewwindow]{http://www.uic.unn.ru/~zny/skeleton/}{\texttt{skeleton}}, and 
\texttt{lcdd\_gmp} from \href[pdfnewwindow]{https://www.inf.ethz.ch/personal/fukudak/cdd\_home}{\texttt{cddlib}}, 
the latter notably being readily available on \href[pdfnewwindow]{https://packages.ubuntu.com/zesty/amd64/libcdd-tools/filelist}{Linux} and 
\href[pdfnewwindow]{https://github.com/Homebrew/homebrew-science}{MacOS}. An especially versatile computational geometry suite for Linux is \href[pdfnewwindow]{https://polymake.org/doku.php/howto/install\#distributions}{\texttt{polymake}}.
}
.

We introduce the notation $\kappa$ as a discrete variable ranging over the vertices, and we indicate the (explicit) noncontextual measurement assignment of vertex $\kappa^*$ by the $(ld)$-component vector ${\tilde{\bm{\xi}}(\kappa^*)\!\equiv\!\Big(\! \tilde{\xi}_{1|M_1}(\kappa^*),\!...,\tilde{\xi}_{d|M_1}(\kappa^*), \tilde{\xi}_{1|M_2}(\kappa^*),\!..., \tilde{\xi}_{d|M_l}(\kappa^*)   \!\Big)}$. %

Now, since any point in a polytope can be written as a convex mixture of the vertices, the noncontextual measurement-assignment polytope can be defined alternatively but equivalently as the convex hull of its vertices:
\setlength{\belowdisplayskip}{0pt plus 0pt}
\setlength{\abovedisplayskip}{0pt plus 0pt}
\setlength\abovedisplayshortskip{0pt}
\setlength\belowdisplayshortskip{0pt}
\begin{samepage}
\begin{thepolytope}\label{measurementpolytopeverts}
The $(ld)$-component vector $\bm{\xi}(\lambda^*)$ lies inside the noncontextual measurement-assignment polytope if and only if there exist some convex weights $\{w(\kappa|\lambda^*)\}_{k}$ such that:
\beq
\forall i,m: \quad \xi_{m|M_i}(\lambda^*) = \sum_{\kappa} w(\kappa|\lambda^*)\tilde{\xi}_{m|M_i}(\kappa) \,,\label{decomp}
\eeq
where $\kappa$ ranges over the vertices found by performing vertex enumeration on the linear constraints of characterization \ref{measurementpolytope}.
\end{thepolytope}
\end{samepage}\normalsize

\noindent Below, we presuppose that one has indeed characterized the noncontextual measurement-assignment polytope by finding its vertices explicitly.%

\subsection{Constructing a noncontextual model with known ontic states and linearly constrained parameters}

Suppose one has a universally noncontextual model of the experiment in the sense of formulation F1, where the ontic state space need not be of finite cardinality. The results of the previous subsection imply that it is always possible to infer from this model another universally noncontextual model wherein the ontic state space is of finite cardinality, as follows.  

By substituting Eq.~\eqref{decomp} into Eq.~\eqref{probs}, each operational probability can be written in terms of a finite sum:
\begin{align}
p(m &| M_i,P_j) = \int_{\Lambda} \xi_{m|M_i}(\lambda) \mu_{P_j}(\lambda) d\lambda \\
&= \int_{\Lambda} \Big( \sum_{\kappa} \tilde{\xi}_{m|M_i}(\kappa)w(\kappa|\lambda) \Big) \mu_{P_j}(\lambda) d\lambda \\
&=  \sum_{\kappa} \tilde{\xi}_{m|M_i}(\kappa) \Big( \int_{\Lambda}w(\kappa|\lambda) \mu_{P_j}(\lambda) d\lambda \Big) \\
&= \sum_{\kappa} \tilde{\xi}_{m|M_i}(\kappa) \nu_{P_j}(\kappa),\label{newprobs}
\intertext{where we have defined}
\label{newdist}
\nu_{P_j}(\kappa) &\equiv \int_{\Lambda}  w(\kappa|\lambda) \mu_{P_j}(\lambda) d\lambda.
\end{align}
Because $\nu_{P_j}(\kappa^*)$ for a given vertex $\kappa^*$ is a convex combination of the values of $\mu_{P_j}(\lambda)$, we can infer that each $\nu_{P_j}(\kappa)$ is a valid probability distribution, and furthermore that the set $\{ \nu_{P_j}(\kappa) \}$ respects noncontextuality with respect to the operational equivalences in $\mathcal{OE}_P$.

Thus, if any noncontextual ontological model exists, then there must also exist a noncontextual model with an ontic state space of finite cardinality. The latter model is constructed by identifying one ontic state with each extremal noncontextual measurement assignment, and then imagining every preparation as a probability distribution over those ontic states, as done in Eq.~\eqref{newdist}. In other words,
\setlength{\belowdisplayskip}{0pt plus 0pt}
\setlength{\abovedisplayskip}{0pt plus 0pt}
\setlength\abovedisplayshortskip{0pt}
\setlength\belowdisplayshortskip{0pt}
\begin{samepage}
\begin{formulation}  \label{c2} 
For a data table $\{ p(m|M_i,P_j) \}_{i,j,m}$, an ontological model that is universally noncontextual with respect to the operational equivalences in $\mathcal{OE}_P$ and $\mathcal{OE}_M$ exists 
 if and only if 
\vspace{1.5pt}\begin{align*}
\exists  \{\nu_{P_j}(\kappa)\}_{j,\kappa} \text{  such that: } \nonumber
\end{align*}\vspace{-2pt}\begin{alignat}{2}
\forall \kappa, j: \quad&& \quad  \nu_{P_j}(\kappa) &\ge 0,  \label{mc0}\\[4pt] 
\forall j: \quad&& \quad \sum_{\kappa} \nu_{P_j}(\kappa) &= 1\,, \label{mc1} \\ 
\forall \kappa, s:\quad&& \quad \sum_j  (\alpha^{s}_{P_j}-\beta^{s}_{P_j}) \nu_{P_j}(\kappa) &= 0\,, \label{mc2} 
\end{alignat}\begin{flalign}
\forall i,j,m:\quad\sum_{\kappa} \tilde{\xi}_{m|M_i}(\kappa) \nu_{P_j}(\kappa) =  p(m | M_i,P_j),  \vspace{4ex}\label{mc3}
\end{flalign}\normalsize
where $\kappa$ ranges over the discrete set of vertices of the polytope defined by Eqs.~\eqref{measurementpolytope} or \eqref{measurementpolytopeverts}.
\end{formulation}
\end{samepage}\normalsize

In this formulation, each operational probability $p(m | M_i,P_j)$ is given as a {\em linear} function of a {\em finite} set of unknown parameters.  This is because the only unknown parameters on the left-hand side of Eq.~\eqref{mc3}
 are $\{ \nu_{P_j}(\kappa)\}_{\kappa}$, while the $\{\tilde{\xi}_{m|M_i}(\kappa)\}_{\kappa}$ are specified numerically---they are the solution of the vertex enumeration problem described in the previous section. Achieving linearity in all the constraints is a critical intermediate step towards finding a final quantifier-free formulation, as we do in the next section, and is critical for the numerical methods we introduce in Section~\ref{satisfy}.

\subsection{The inequalities formulation of the generalized-noncontextual polytope} \label{linearQE}

To obtain constraints that refer only to operational probabilities, we eliminate the unobserved $\{\nu_{P_j}(\kappa)\}_{j,\kappa}$ from the system of equations~\eqref{c2}, obtaining a system of linear inequalities over the $\{p(m | M_i,P_j)\}_{i,j,m}$ alone. The linearity of the final inequalities follows from the linearity of the inequalities and equalities in 
Eqs.~\eqref{c2}. This establishes that the space of noncontextual data tables defines a polytope.

The standard method for solving this problem of \textit{linear quantifier elimination} is the Chernikov-refined Fourier-Motzkin algorithm~\cite{DantzigEaves,BalasProjectionCone,jones2004equality,Shapot2012,Bastrakov2015}, which is implemented in a variety of software packages\footnote{Dedicated software for eliminating variables from a set of linear inequalities via the Fourier-Motzkin algorithm includes \texttt{fmel} from \href[pdfnewwindow]{http://porta.zib.de}{\texttt{PORTA}}, \texttt{fme} from \href[pdfnewwindow]{http://sbastrakov.github.io/qskeleton/}{\texttt{qskeleton}}, and \texttt{fourier} from \href[pdfnewwindow]{http://cgm.cs.mcgill.ca/~avis/C/lrslib/USERGUIDE.html\#fourier}{\texttt{lrs}}. From a geometric perspective, each variable in a linear system is an axis of some high-dimensional coordinate system; consequently, eliminating a variable is equivalent to projecting the polytope onto a hyperplane orthogonal to that particular axis. As such, polytope projection is the titular topic of most of the relevant literature on linear quantifier elimination~\cite{DantzigEaves,BalasProjectionCone,jones2004equality,Shapot2012,Bastrakov2015}. Furthermore, polytope projection and vertex enumeration are intimately related: One can define the vertex enumeration problem as a task of linear quantifier elimination, and therefore any polytope projection algorithm can be used to perform vertex enumeration, albeit less efficiently than specialized algorithms~\cite{Avis2000lrs,bremner_symmetries_2009,Zolotykh2012,avis_convexhull_2015,panda_2015}. Conversely, a brute-force technique for polytope projection is to first enumerate the polytope's vertices, manually discard the to-be-eliminated coordinates from each each vertex, and then reconvert back to inequalities using a convex hull algorithm. This roundabout method of performing polytope projection is generally suboptimal, but can be used in practice.}. 

Denoting the quantifier-free list of linear facet inequalities of the generalized-noncontextual polytope by $\{h^1, h^2,...,h^n\}\equiv\mathcal{H}$ (for `halfspaces') and letting the coefficients of a specific facet inequality $h$ be given by $\gamma^h_{i,j,m}$ while $\gamma^h_0$  indicates the constant term in that inequality, we find that:
\begin{samepage}
\begin{formulation}\label{c3}
For a data table $\{ p(m|M_i,P_j) \}_{i,j,m}$, an ontological model that is universally noncontextual with respect to the operational equivalences in $\mathcal{OE}_P$ and $\mathcal{OE}_M$ exists %
 if and only if
\begin{align} %
\forall h\in\mathcal{H}\,:\quad\sum_{i,j,m} \gamma^h_{i,j,m} p(m|M_i,P_j) + \gamma^h_0 \ge 0\,,
\end{align}
where $\mathcal{H}$ is the set of $n$ inequalities resulting from eliminating all free parameters $\{\nu_{P_j}(\kappa)\}_{j,\kappa}$ in the formulation of Eq.~\eqref{c2}. 
\end{formulation}\end{samepage}

\section{Does a given numerical data table admit of a noncontextual model?} \label{satisfy}

To date, experimental tests of generalized noncontextuality~\cite{POM,unwarranted} have targeted the \textit{specific} preparations, measurements, and operational equivalences of some particular quantum no-go theorem~\cite{peresmermin,POM,unwarranted,stated,statei,deba2}. By abstracting away the quantum-specific elements of the proof and describing the experiment in entirely operational terms, one can identify operational features of a set of preparations and measurements, such that any theory exhibiting those features fails to admit of a noncontextual ontological model. To test these operational features, previous experiments have used post-processed data to enforce specific operational equivalences appearing in the quantum no-go argument. (See the ``secondary procedures'' technique described in Ref.~\cite{unwarranted}.)

We here introduce a much more general analysis technique, in which one need not target any specific preparations, measurements, and operational equivalences. Rather, arbitrary numerical data tables can be directly analyzed. With respect to \textit{whatever} operational equivalences happen to be manifest in the data (see Footnote~\ref{foot1}), one can use the methods we present below to efficiently test whether the numerical data table admits of a noncontextual model or not. Because answering this yes-no question does not require deriving the full set of noncontextuality inequalities for the scenario under study, it is computationally very efficient\footnote{An analagous pair of problems with widely differing computational difficulties has long been appreciated it the study of Bell nonlocality. Obtaining \emph{all}  the Bell inequalities which characterize some nonlocality scenario can be quite difficult, but ascertaining if a particular correlation admits a local model or not can be resolved with the application of a single linear program~\cite{Zukowski1999,Basoalto2001}. The same (efficient) linear program can be used to return a \emph{single} Bell inequality which certifies the nonlocality of the given correlation~\cite{Elliot2009}.}. Furthermore, analyzing data in this manner always allows for larger inequality violations, since the post-processing required in the secondary procedures technique of Ref.~\cite{unwarranted} always introduces additional noise.

To test whether a numerically specified data table $\{ p^*(m|M_i,P_j) \}_{i,j,m}$ admits of a noncontextual model, we leverage the formulation in Eq.~\eqref{c2}. All the equality constraints of Eqs.~\eqref{mc1}, \eqref{mc2}, and \eqref{mc3} can be encoded in a single matrix equality constraint, 
\beq
\bf{M}\cdot \bf{x} = \bf{b}^*,
\eeq
where $\bf{M}$ contains the parameters $\alpha^{s}_{P_j} - \beta^{s}_{P_j}$ and the quantities $\{ \tilde{\xi}_{m|M_i}(\kappa) \}_{i,m,\kappa}$, $\bf{x}$ contains the unknown parameters $\{\nu_{P_j}(\kappa)\}_{j,\kappa}$, and $\bf{b}^*$ contains the probabilities $\{ p^*(m|M_i,P_j) \}_{i,j,m}$, as well as zeroes and ones corresponding to the right-hand sides of Eqs.~\eqref{mc1} and \eqref{mc2}.  Eq.~\eqref{mc0} becomes simply ${\bf x} \ge 0$. Hence, for a numerically specified $\{ p^*(m|M_i,P_j) \}_{i,j,m}$, the formulation of Eq.~\eqref{c2} defines a linear program (LP)\footnote{Linear programming is used across many fields; specialized LP software packages include \href[pdfnewwindow]{https://www.mosek.com/resources/downloads}{\textit{Mosek}}, \href[pdfnewwindow]{http://www.gurobi.com/products/gurobi-optimizer}{\textit{Gurobi}}, and \href[pdfnewwindow]{https://www-01.ibm.com/software/commerce/optimization/linear-programming/}{\textit{CPLEX}}.}. 

The primal LP is the search for a solution to a linear system of equations, namely:
\begin{flalign}\begin{split}\label{eq:primalexist}
\exists \ \bf{x}&\text{ such that}\\
\bf{M}\cdot \bf{x} &= \bf{b}^*\,,\\
\text{and }\:\bf{x} &\geq \bf{0}\,.
\end{split}\end{flalign}
Because no objective function to maximize or minimize is specified in the LP defined by Eq.~\eqref{eq:primalexist}, this means the LP is just checking for the existence of an $\bf{x}$ which satisfies the constraints and hence guarantees the existence of a noncontextual model, via Eq.~\eqref{c2}. 

Whenever the primal LP is infeasible -- that is, no solution can be found -- one can obtain a \textit{certificate of primal infeasibility}, also known as the \textit{Farkas dual}~\cite{infeasibilitycertificates,YinyuYe2016LP}. The certificate of primal infeasibility is obtained by solving the complementary\footnote{The complementary LP defined in Eq.~\eqref{eq:dualminimize} is meant to explain how infeasibility certificates are generated in practice. Note, however, that the Farkas dual of an LP is not the same as the LP's dual formulation, although the concepts are related. See Refs.~\cite{infeasibilitycertificates,YinyuYe2016LP}, as well as Ref.~\cite[Theorem 1 and supplementary materials]{NCFraction}).} %
 linear system
\begin{flalign}\begin{split}\label{eq:dualminimize}
\min_{\bf{y}} \quad \bf{y}\cdot \bf{b}^*  &\text{ such that}\\
\bf{1}\ge \bf{y}\cdot\bf{M} &\geq \bf{0}\,.
\end{split}\end{flalign}
Farkas' lemma states that either the primal LP is feasible, or else the certificate $\bf{y}$ resulting from Eq.~\eqref{eq:dualminimize} satisfies the strict inequality $\bf{y}\cdot \bf{b}^*<0$.

Farkas's lemma is easily proven: Plainly, if there exists such a $\bf{y}$ (i.e, not only $\bf{y}\cdot\bf{M} \ge \bf{0}$ but also $\bf{y}\cdot \bf{b}^*<0$), then there \emph{cannot} exist an $\bf{x}$ which satisfies the primal LP of Eq.~\eqref{eq:primalexist}; since the inequalities
\begin{subequations}
\begin{alignat}{2}
&&\bf{x} &\ge \bf{0}\,,\\
\text{and }\:&&\bf{y}\cdot\bf{M} &\ge \bf{0}\,,\\
\text{and }\:&&\bf{y}\cdot\bf{b^*} = \bf{y}\cdot\bf{M}\cdot \bf{x} &< 0\,,
\end{alignat}\end{subequations}
can not all be satisfied simultaneously.

Of relevance to this work is that we may interpret the certificate $\bf{y}$ resulting from Eq.~\eqref{eq:dualminimize} as a noncontextuality inequality, since Farkas' lemma ensures that $\bf{y}\cdot\bf{b}\ge~0$ for every $\bf{b}$ for which the primal LP is feasible. The extent to which $\bf{y}\cdot \bf{b}^*$ is negative is identically the amount by which the corresponding noncontextuality inequality is violated by the (contextual) $\{p^*(m|M_i,P_j)\}_{i,j,m}$. When interpreting certificates of primal infeasibility as noncontextuality inequalities, one deduces the constant term from those elements of $\bf{b^*}$ which do not depend on $\{ p^*(m|M_i,P_j) \}_{i,j,m}$. In practice, therefore, the constant term is the sum of those elements in $\bf{y}$ which correspond to the normalization conditions of Eq.~\eqref{mc1}.

This technique allows an experimenter to optimally certify the contextuality of a numerical data table $\{ p^*(m|M_i,P_j) \}_{i,j,m}$ without first performing the computationally expensive task of finding all the noncontextuality inequalities, i.e., without doing any work to transform formulation \ref{c2} into formulation \ref{c3}. Instead, by seeking a certificate of primal infeasibility -- a single query to a linear program -- one obtains \emph{the} noncontextuality inequality which best witnesses the contextuality of the data table. 

\section{Applications}
\subsection{Generalized-noncontextual polytope in the simplest nontrivial case} \label{sec:simple}
As argued in Ref.~\cite{robust}, the simplest possible scenario in which the principle of noncontextuality implies nontrivial constraints on operational probabilities involves four preparations and two binary-outcome measurements\footnote{Ref.~\cite{robust} also assumes that these two measurements are tomographically complete, but we do not make this assumption here. See Refs.~\cite{robust,unwarranted} for details on the issue of tomographic completeness.}. We imagine for simplicity that the preparations satisfy the operational equivalence 
\beq \label{OE1}
\frac{1}{2}P_1+\frac{1}{2}P_2 \simeq \frac{1}{2}P_3+\frac{1}{2}P_4
\eeq
and that there are no operational equivalences among the measurements.

We denote the operational probability $p(0|M_i,P_j)$ by $p_{ij}$. (By normalization, probability $p(1|M_i,P_j)$ is then $1- p_{ij}$.) Vertex enumeration finds 4 vertices for the noncontextual measurement-assignment polytope, corresponding to the four deterministic assignments $(\xi_{0|M_1}(\lambda), \xi_{0|M_2}(\lambda)) \in \{ (0,\!0),(0,\!1),(1,\!0),(1,\!1) \}$. Each of the 4 preparations defines a probability distribution over these 4 ontic states, so there are 16 free parameters to be eliminated. Linear quantifier elimination finds the polytope of data tables consistent with the principle of noncontextuality and the operational equivalence of Eq.~\eqref{OE1} to be:
\begin{subequations}\label{poly}\begin{align}
&\forall i,j:\qquad 0 \le p_{ij} \le 1 \,, \\   
&p_{12}+p_{22}-p_{23}-p_{14} \le 1 \,,\label{eq:polyviolated1}\\ 
&p_{12}+p_{22}-p_{13}-p_{24} \le 1 \,,\\ 
&p_{22}+p_{13}-p_{12}-p_{24} \le 1 \,,\\ 
&p_{12}+p_{23}-p_{22}-p_{14} \le 1 \,,\\ 
&p_{22}+p_{14}-p_{12}-p_{23} \le 1 \,,\\ 
&p_{23}+p_{14}-p_{21}-p_{22} \le 1 \,,\\ 
&p_{12}+p_{24}-p_{22}-p_{13} \le 1 \,,\\ 
&p_{13}+p_{24}-p_{12}-p_{22} \le 1 \,\label{eq:polyviolated}.
\end{align}\end{subequations}
Note that the two probabilities which do not appear in Eqs.~\eqref{eq:polyviolated1}-\eqref{eq:polyviolated}, $p_{11}$ and $p_{21}$,
 are fixed by the operational equivalence relation, Eq.~\eqref{OE1}:
\begin{subequations}\label{eq:impliciteqs}\begin{align}
&p_{11} = p_{13}+p_{14}-p_{12} \,,\\ 
&p_{21} = p_{23}+p_{24}-p_{22} \,.
\end{align}\end{subequations}

Ineqs.~\eqref{poly} and Eqs.~\eqref{eq:impliciteqs} tightly define the generalized-noncontextual polytope for this scenario. Furthermore, all of Ineqs.~\eqref{poly} are equivalent under relabeling. That is, any one of the inequalities can generate all 8 by applying relabelings which respect the operational equivalences: $M_1 \leftrightarrow M_2$, $P_1 \leftrightarrow P_2$, and $(P_1,P_2) \leftrightarrow (P_3,P_4)$. Similarly, each of Eqs.~\eqref{eq:impliciteqs} is equivalent to the other under the same relabelings.

As an illustration of how a noncontextuality inequality can be derived
from a numerically specified (contextual) data table, consider the following example
\begin{align}\begin{split}\label{eq:contextualdata}
p_{11}= 1,\quad p_{12}= 0,\quad p_{13}=
   1,\quad p_{14}= 0,\\
   p_{21}= 1,\quad p_{22}=
   0,\quad p_{23}= 0,\quad p_{24}= 1,
\end{split}\end{align}
which respects the operational equivalence relation of Eq.~\eqref{OE1}, but maximally violates Ineq.~\eqref{eq:polyviolated}, since it has ${p_{13}+p_{24}-p_{12}-p_{22}=~2\not\leq 1}$. 
Indeed, when we construct the primal linear program per Eq.~\eqref{eq:primalexist}, we find it to be infeasible, and we find that the certificate of infeasibility returned by our numerical solver corresponds to Ineq.~\eqref{eq:polyviolated}. %

\subsubsection{Relevance to parity-oblivious multiplexing} \label{sec:POMP}

In the communication task of ``parity-oblivious multiplexing'', an agent Alice wishes to communicate two bits to an agent Bob, in such a way that Bob can extract information about either of the two bits but cannot extract any information about their parity~\cite{POM}.  This task involves four preparations (associated to the four possibilities for the values of the two bits) and two measurements (corresponding to which bit Bob wishes to learn about), and the parity-obliviousness condition implies an operational equivalence relation among the preparations, namely, that of Eq.~\eqref{OE1}.  Consequently, this task fits precisely the 
operational scenario considered in this section.
 
 In Ref.~\cite{POM}, it was shown that contextuality provides an advantage for the task of parity-oblivious multiplexing: the maximum probability of succeeding at this task in a noncontextual model is $\nicefrac{3}{4}$, so that any higher probability of success requires contextuality. In a quantum world, for instance, one can succeed with probability $\frac{1}{4}(2+\sqrt{2}) \approx 0.85$.

If one identifies our preparations $P_1$, $P_2$, $P_3$, and $P_4$ with Ref.~\cite{POM}'s preparations $P_{00}$, $P_{11}$, $P_{01}$, and $P_{10}$, respectively, then the inequality in Ref.~\cite{POM} is our facet Ineq.~\eqref{eq:polyviolated}. This is the same inequality which witnesses the failure of noncontextuality for the data table defined in Eq.~\eqref{eq:contextualdata}, which is to be expected, as this particular data table describes a set of probabilities which achieve the maximum logically possible probability of success at parity-oblivious multiplexing.

Additionally, one could apply our algorithm to generalized types of parity-oblivious multiplexing. For instance, Ref.~\cite{POM} derived a bound on the probability of success in a noncontextual model of $n$-bit parity-oblivious multiplexing, in which Alice wishes to communicate $n$ bits to Bob under the constraint that Bob can learn no information about the parity between any two of the bits. These bounds are tight, and can be saturated by a na\"{i}ve classical strategy. However, one could further use our techniques to learn whether or not they constitute facet inequalities of the generalized-noncontextual polytope, as well as to find the full generalized-noncontextual polytope. 

For the still more general case where the different $n$-bit strings which Alice wishes to send do not have equal {\em a priori} probabilities, our method can also find the generalized-noncontextual polytope, from which one can immediately infer the maximum success probability for the task.

\subsection{Generalized-noncontextual polytopes for scenarios relevant to state discrimination} \label{sec:MESD}

One way in which one can generalize the simplest operational scenario described above is to increase the number of binary-outcome measurements from two to three while still not assuming any operational equivalences among them, so that the only operational equivalence relation remains the one between the preparations (Eq.~\eqref{OE1}).  This operational scenario can also be related to an information-theoretic task, namely, the task of minimum error state discrimination, as noted by  two of the present authors in Ref.~\cite{MESD}. 

In the quantum version of this task, an agent wishes to guess which of two pure quantum states a system was prepared in
given a single sample of the system, where the identity of the two quantum states is known.  
Quantum theory prescribes a particular trade-off relation between the probability of success and the non-orthogonality of the two quantum states, and Ref.~\cite{MESD} showed that this trade-off contradicts the principle of generalized noncontextuality.  The ideal quantum realization of minimum error state discrimination fits the operational scenario described above: the two pure quantum states define two of the preparation procedures, while their orthogonal complements in the 2d subspace that they span define the other two.  The fact that the equal mixture of any two orthogonal pure states in a 2d subspace is independent of the basis implies the operational equivalence of Eq.~\eqref{OE1}.  Finally, the degree of nonorthogonality has an operational interpretation as the probability of one state passing a test for the other (termed the {\em confusability}).  Therefore, the measurements of each of the two bases, together with the discriminating measurement, provide the three binary-outcome measurements in the scenario.

 The facet noncontextuality inequalities for this operational scenario are given in Appendix D of Ref.~\cite{MESD}~\footnote{Actually, the polytope given therein is the intersection of the generalized-noncontextual polytope with two additional inequalities, which are implied by making sensible labeling choices.}, and these are seen to imply a nontrivial upper bound on the probability of successful discrimination for a given confusability.  
Hence, contextuality provides an advantage for minimum error state discrimination. The quantum probability of successful discrimination for a given confusability 
is higher than that allowed in a noncontextual model, and hence partakes in this contextual advantage.

Using our technique, one can also immediately derive the generalized-noncontextual polytope for more general minimum error state discrimination scenarios, such as those in which the quantum states (preparations) being discriminated are sampled with unequal probabilities, or in which there are more than two quantum states (preparations).

Because state discrimination in various forms is a primitive for many other quantum information-processing tasks, such analyses should be valuable for identifying the circumstances in which contextuality constitutes a resource.

\subsection{Generalized-noncontextual polytopes for a scenario involving both preparation and measurement noncontextuality} \label{sec:coin} 

So far, our examples have involved operational equivalences only among preparations. In this section, we revisit the scenario considered in the recent experimental test of noncontextuality in Ref.~\cite{unwarranted}, which involves operational equivalences among the preparations and also among the measurements. Specifically, we imagine a set of six preparations and three binary-outcome measurements, where the preparations satisfy the operational equivalences 
\beq \label{OE2}
\frac{1}{2}P_1+\frac{1}{2}P_2 \simeq \frac{1}{2}P_3+\frac{1}{2}P_4 \simeq \frac{1}{2}P_5+\frac{1}{2}P_6,
\eeq
and the measurement effects satisfy the operational equivalence
\begin{align}
\begin{aligned} \label{OE3}
\frac{1}{3}[0|M_1]&+\frac{1}{3}[0|M_2]+\frac{1}{3}[0|M_3] \\
\simeq &\frac{1}{3}[1|M_1]+\frac{1}{3}[1|M_2]+\frac{1}{3}[1|M_3].
\end{aligned}
\end{align}
(See Ref.~\cite{unwarranted} for a discussion of the significance of these operational equivalences.)

We denote the operational probability $p(0|M_i,P_j)$ by $p_{ij}$ and $p(1|M_i,P_j)$ by $\overline{p}_{ij}$. Vertex enumeration finds 6 vertices for the noncontextual measurement-assignment polytope, corresponding to the four \textit{in}deterministic assignments defined by $(\xi_{0|M_1}(\lambda), \xi_{0|M_2}(\lambda),\xi_{0|M_3}(\lambda)) \in \{ (0,\!\frac{1}{2},\!1),(\frac{1}{2},\!0,\!1),(1,\!0,\!\frac{1}{2}),(1,\!\frac{1}{2},\!0),(0,\!1,\!\frac{1}{2}),(\frac{1}{2},\!1,\!0) \}$. Each of the 6 preparations defines a probability distribution over these 6 ontic states, so there are 36 free parameters to be eliminated. Linear quantifier elimination finds the polytope of data tables consistent with the principle of noncontextuality and with the operational equivalences of Eq.~\eqref{OE2} and Eq.~\eqref{OE3}. We find that this polytope has 1596 facet inequalities. 

Plainly, 1596 inequalities is far too many to list explicitly. %
However, by considering the physical symmetries of this scenario, we can significantly simplify our description of these facets. Since the scenario is invariant under various relabelings of measurements [Eqs.~(\ref{gen:mswap1}-\ref{gen:mswap2})], outcomes [Eq.~\eqref{gen:moutcomeswap}], and preparations [Eqs.~(\ref{gen:pswapinpair}-\ref{gen:sourceswap2})]---i.e. those relabelings which respect the operational equivalences---we know a priori that the generalized-noncontextual polytope will possess significant internal symmetry. The symmetry group which leaves our polytope invariant is generated by the six relabelings
\begin{subequations}\begin{align}
M_1 &\leftrightarrow M_2\label{gen:mswap1}\\
M_1 &\leftrightarrow M_3\label{gen:mswap2}\\
([0|M_1],[0|M_2],[0|M_3]) &\leftrightarrow ([1|M_1],[1|M_2],1|M_3])\label{gen:moutcomeswap}\\
P_1 &\leftrightarrow P_2\label{gen:pswapinpair}\\
(P_1,P_2) &\leftrightarrow (P_3,P_4)\label{gen:sourceswap1}\\
(P_1,P_2) &\leftrightarrow (P_5,P_6)\label{gen:sourceswap2}.
\end{align}\end{subequations}
We use parentheses to indicate a coherent relabelling: e.g., the outcomes of three measurements can be flipped per Eq.~\eqref{gen:moutcomeswap}, but only if all three measurements have their outcomes relabeled simultaneously. This is in contrast to an exchange like $P_1 \leftrightarrow P_2$ per Eq.~\eqref{gen:pswapinpair}, which can be performed in isolation. The total order of this symmetry group is 576.

Under this group, we find that the 1596 facet inequalities admit classification into seven symmetry classes. We therefore explicitly list a single representative inequality from each class:
\begin{align}\label{poly2}
\hspace{-2ex}\begin{tabular}{rrrrrrl||r}
 \multicolumn{6}{l}{Inequality Terms}  & \(\substack{\text{Upper}\\ \text{Bound}\\{}}\!\) & \(\substack{\text{Orbit}\\ \text{Size}\\{}}\!\) \\
\hline
 \(p_{11}\!\) &  &  &  &  &  & $\;\leq\;$1 & \(36\!\) \\
 \(p_{11}\!\) &  &  &  & \(+p_{23}\!\) & \(+p_{35}\!\) & $\;\leq\;$\(2.5\) & \(48\!\) \\
 \(p_{11}\!\) &  &  & \(+p_{22}\!\) &  & \(+p_{35}\!\) & $\;\leq\;$\(2.5\) & \(72\!\) \\
 \(p_{11}\!\) & \(-p_{13}\!\) &  \(-2\, p_{15}\!\) & \(-2\, p_{22}\!\) &  \(+2\, p_{23}\!\) & \(+2\, p_{35}\!\) & $\;\leq\;$3 & \(576\!\) \\
 \(2\, p_{11}\!\) &  &  & \(-p_{22}\!\) & \(+2\, p_{23}\!\) &  & $\;\leq\;$3 & \(144\!\) \\
 \(p_{11}\!\) &  & \(-p_{15}\!\) & \(+p_{22}\!\) & \(+p_{23}\!\) & \(+2\, p_{35}\!\) & $\;\leq\;$4 & \(576\!\) \\
 \(p_{11}\!\) &  & \(-p_{15}\!\) & \(+2\, p_{22}\!\) &  & \(+2\, p_{35}\!\) & $\;\leq\;$4 & \(144\!\) \\
\end{tabular}\end{align}
The number of inequalities in each symmetry class is given by the ``Orbit Size" in Ineqs.~\eqref{poly2}. 
The generalized-noncontextual polytope for this scenario is defined by the 1596 facet inequalities, as well as by equalities which hold for \emph{any} data table (contextual or noncontextual) admitting the operational equivalence relations per Eqs.~\eqref{OE2} and \eqref{OE3}. These equalities fall into three distinct symmetry classes, represented by 
\begin{subequations}\label{eq:impliciteqs2}\begin{align}
&p_{11} + p_{14} = p_{12}+p_{15} \,\\ 
&p_{11} + p_{21} + p_{31} = 3/2 \,\\
&p_{11}+\overline{p}_{11}=1.
\end{align}\end{subequations}
The first two equalities are enforced by the operational equivalence relations, Eqs.~\eqref{OE2} and \eqref{OE3}, respectively, while the third equality is guaranteed by normalization of measurements, Eq.~\eqref{m2}.

To test if a given data table lies inside this generalized-noncontextual polytope, one could reconstruct all 1596 inequalities from the seven given in Ineqs.~\eqref{poly2}, but it is likely much easier to instead artificially generate equivalent-up-to-symmetries data tables from one's actual data table, and then to test each of \emph{those} against the seven canonical inequalities\footnote{The first inequality in Ineqs.~\eqref{poly2} is inviolable, even by contextual data tables, so technically one only needs to test against the remaining six inequalities.}. One only needs to consider at most 576 data table variants (per the group order), although in practice there will be fewer variants to consider if the data table one wishes to investigate possesses any internal symmetry of its own.

Noting that\footnote{Eq.~\eqref{eq:relatingtoformer} is a nontrivial but readily verifiable consequence of the inviolable equalities in Eqs.~\eqref{eq:impliciteqs2}.}
\beq\label{eq:relatingtoformer}
p_{11}+p_{23}+p_{35}=\overline{p}_{12}+\overline{p}_{24}+\overline{p}_{36}\,,
\eeq
the single inequality derived (and experimentally violated) in Ref.~\cite[Eq.~(6)]{unwarranted} is recognized as a facet of the generalized-noncontextual polytope, namely it is precisely the second inequality in Ineqs.~\eqref{poly2}:
\beq \label{eq:polyviolated2}
2\left( p_{11}+ p_{23}+ p_{35}\right)\le 5\,.
\eeq
One can also derive Ineq.~\eqref{eq:polyviolated2} directly (and efficiently!) by using the linear program presented in Section~\ref{satisfy}. Namely, it is the inequality corresponding to the certificate of infeasibility returned by our numerical solver, when we construct the primal linear program using Eq.~\eqref{eq:primalexist} together with the ideal (contextual) quantum data table.

\section{Connections to Bell polytopes and to polytopes of general probabilistic models in Kochen-Specker contextuality scenarios}%

The central object of interest in prepare-and-measure contextuality scenarios are data tables, in particular the polytope of noncontextual data tables discussed herein. Along the way to finding this polytope, we have found it useful to first compute the  noncontextual measurement-assignment polytope. It turns out that any Bell polytope is directly connected to our generalized-noncontextual polytope for a corresponding prepare-and-measure scenario, and that the polytope of generalized probabilistic models of Ref.~\cite{AFLS} (and instances thereof, such as the polytope of `no-disturbance correlations' defined in Ref.~\cite{ncycle}) are special cases of our noncontextual measurement-assignment polytope.

In the literature on Bell inequalities, the set of allowed classical correlations is known as the {\em Bell polytope} (or equivalently the local polytope)~\cite{Fine, Pitowsky}. As discussed in Ref.~\cite{LSW}, any Bell scenario can be analyzed as a prepare-and-measure contextuality experiment. 
In brief, one can always reimagine the measurements of one of the parties in the Bell scenario as a remote preparation of the distant party's state (via quantum steering). Under this mapping, the no-signaling condition in the Bell scenario implies an operational equivalence in the contextuality scenario: that the average state of the steered system is the same for any choice of steering measurement. Further, under this mapping the assumption of local causality implies the assumption of preparation noncontextuality for this operational equivalence relation. 
Then, one finds that our generalized-noncontextual polytope is directly connected to the Bell polytope: specifically, the latter is generated from the former by multiplying each prepare-and-measure probability by the probability of steering to the corresponding remote preparation. %

For example, our contextuality scenario in Section~\ref{sec:simple} is isomorphic to the Clauser-Horne-Shimony-Holt Bell scenario~\cite{CHSH} (with two parties, each with two binary outcome measurements), as discussed in Ref.~\cite{POM}. Additionally, our contextuality scenario in Section~\ref{sec:MESD} is isomorphic to a Bell scenario (with two parties, one with two binary measurements and another with two ternary measurements), as discussed in Ref.~\cite{MESD}. %

In this way, all Bell scenarios are special cases of our framework. However, our framework is much more general. As stated in Ref.~\cite{MESD}, the only prepare-and-measure contextuality scenarios which are equivalent to a related Bell scenario are those which do not leverage any operational equivalences among the measurements, and which leverage only those operational equivalences among the preparations which arise from the various ensemble decompositions of a single mixed preparation.

There is also a connection between our work and that of Ref.~\cite{AFLS}, which gives an operational framework for studying Kochen-Specker contextuality scenarios. Ref.~\cite{AFLS} considers an arbitrary set of measurements, some of which have one or more measurement effects in common, and characterizes various classes of probability assignments to those measurement effects (i.e., various classes of what we have here called ``measurement assignments''). They term these classes of assignments `classical', `quantum', `general probabilistic', etc. 
Note, however, that the class of assignments termed `classical'
in Ref.~\cite{AFLS}, does {\em not} correspond to the noncontextual measurement-assignment polytope in our approach, since Ref.~\cite{AFLS} assumes outcome determinism, while in our approach {\em indeterministic} probability assignments are allowed.
Rather, it is the polytope of {\em general} probabilistic models in Ref.~\cite{AFLS} that corresponds to our noncontextual measurement-assignment polytope because the definition of a general probabilistic model 
for the scenarios considered
 in Ref.~\cite{AFLS} 
 is precisely our definition of a measurement-noncontextual assignment (which, we recall, can be indeterministic).
 The two polytopes coincide exactly in every scenario to which the framework of Ref.~\cite{AFLS} applies, since in that case the operational equivalences among the measurements are precisely those given by the occurrences of some effect in more than one measurement. However, our noncontextual measurement-assignment polytope is also defined for scenarios with other types of operational equivalences among the measurements, where the framework of Ref.~\cite{AFLS} does not apply. 

\section{Conclusions}

For arbitrary prepare-and-measure experiments, we have presented a method for finding necessary and sufficient conditions for a data table to admit of a noncontextual model, subject to any fixed sets of operational equivalences among preparations and among measurements. We have also presented an efficient method for determining whether a \textit{numerical} data table is noncontextual, in this same setting. 

We have provided worked examples of each of these methods, in the process deriving necessary and sufficient conditions for operational scenarios in which only necessary conditions were previously known. Equivalently, we have derived the full generalized-noncontextual polytopes for scenarios in which only a single facet inequality was previously known. The operational scenario studied in Section~\ref{sec:simple} is of relevance to parity-oblivious multiplexing~\cite{POM}, while the operational scenario studied in Section~\ref{sec:coin} originates in a recent experimental test of contextuality~\cite{unwarranted}.

 A precursor to the current work can be found in~\cite{Anirudh}. A distinct method, introduced in Ref.~\cite{peresmermin}, also allows one to derive all of the facet inequalities for many operational scenarios. However, the method therein is not fully general, as it applies only to scenarios in which one special equivalence class of preparations is singled out (see Section III. B. of Ref.~\cite{peresmermin} for details). It would be interesting to compare the two approaches, e.g., in terms of computational efficiency, and to modify the approach in Ref.~\cite{peresmermin} to make it as general as the approach described in this article.

\begin{acknowledgments}
The authors are grateful to Anubhav Chaturvedi for identifying a typographic error in the inequalities~\eqref{poly2}; the error has been corrected here. D.S. thanks Ravi Kunjwal for useful discussions. This research was supported by a Discovery grant of the Natural Sciences and Engineering Research Council of Canada and by Perimeter Institute for Theoretical Physics. Research at Perimeter Institute is supported by the Government of Canada through the Department of Innovation, Science and Economic Development Canada and by the Province of Ontario through the Ministry of Research, Innovation and Science. 
\end{acknowledgments}

\setlength{\bibsep}{5pt plus 4pt minus 2pt}
\bibliographystyle{apsrev4-2-wolfe}
\nocite{apsrev41Control}
\bibliography{bib}

\end{document}